\documentstyle{paper}
\def\etal    {{ et~al.}~}

\def\amin{$^\prime$}
\def\eg{{\it e.g.}~}

\def\msun     {$M_{\odot}$}
\def\re {\par}

\begin{document}

\title{A Comparison of the ISM in Ellipticals to the Hot Gas in 
a Dense Group from Combined ROSAT and ASCA Observations}

\author{A.Finoguenov, \\
{\it Space Research Institute, Profsoyuznaya 84/32, 117810 Moscow, Russia}\\
W.Forman and C.Jones\\
{\it Harvard-Smithsonian Center for Astrophysics, 60 Garden st., MS 2,
Cambridge, MA 02138, USA}} 

\maketitle

\section*{Abstract}
We analyze diffuse X-ray emission from NGC4649, NGC5044 and NGC5846,
combining data from two X-ray observatories, ROSAT and ASCA.  With ASCA, we
perform a detailed analysis of the X-ray emission which properly accounts
for the ASCA PSF and also include a 3-dimensional source model.  All three
sources exhibit cooling flows in their centers. From the derived abundances
of Si, S and Fe, we conclude that NGC5044 (a dense group of galaxies) was
able to retain significant amounts of gas during an early galaxy wind phase.
Fe abundance in NGC5846 (normal E galaxy) corresponds to current SNIa rate
of $\sim0.2$ of Tammann rate, while low Fe mass in NGC4649 (normal E
in the Virgo cluster) could be an environmental effect.

\section{Data reduction}

To extract the surface brightness profiles we use ROSAT/PSPC data in the
{\mbox{0.5--2.0~keV}} band, omitting point sources. For NGC5044 we also
exclude a region 2\amin\ south-south-east of the center, where a surface
brightness enhancement was detected previously (D94). To calculate ASCA/SIS
ARF matrix, we make a 3-dimensional model of the source based on results of
ROSAT profile analysis, using a $\beta$-model, and assuming spherical
symmetry. For NGC5846 and NGC4649 the second component of the beta fits is
significant. Details are described in FFJ96, where it is also shown that our
procedure is not affected by variation of ROSAT/PSPC count rate with
temperature.  Analysis of the hardness ratio shows, except for the NGC5044
feature noted, that the temperature distribution is symmetrical around the
center.  For the spectral fitting of ROSAT data, we used the
{\mbox{0.2--2.0~keV}} band and for the ASCA/SIS data {\mbox{0.6--2.2~keV}}.
Absorption was taken as the Galactic value. The MEKAL model (from XSPEC-9.0
package) was used in the spectral modeling with abundances from AG89.  The
spectrum of the core region ($<$1\amin) is complex due to the presence of
cooling gas, so we present only the temperature.

\section{Results and discussion}

High abundances of Si, S and Fe in NGC5044 group compared to the two
ellipticals, suggest that more of the heavily enriched gas, produced at
early epochs, was bound by the deeper group potential than by that of the
individual galaxies. A total mass of Si, S and Fe, inside 300 kpc from
NGC5044 center is 0.9 (0.4--1.6), 0.5 (0.2--0.9), 2.9 (2.2--4.7) of $10^{8}$
\msun, respectively with total gas mass of $6.2\pm0.4\times10^{11}$\msun. 
The mass ratio of Fe to Si is larger then predicted on the onset of the
galactic wind (\eg MT87) and the data favors presence of the ``outflow
phase'' in the galactic history.  Abundances of Si, S and Fe, observed in
NGC4649 and NGC5846, are determined by SNIa output and stellar mass loss.
While SNIa supply mainly Fe, a ratio of Si to Fe in stars at the radius
involved is not well known. So far, in order to explain the very presence of
the optical abundance gradients a shorter time of star formation with
increasing radii is assumed (\eg CDB93), one can argue, that the impact from
SNII explosions from more massive progenitors prevail at the outer radius,
which is characterized by higher Si and S to Fe ratio (\eg WW86).
Nevertheless, an upper limit on SNIa output can be set by ascribing all the
observed Fe to the SNIa. Taking SNIa rate slope s=1.4 and initial time of
the ``inflow phase'' equal to 9~Gyr (KRM model in C91), an observed
Fe/Fe$_{\odot}$ abundance value of 0.5 corresponds to current SN Ia rate
values of 0.2 of Tammann rate (for details see FFJ96).  Low mass of Fe,
observed in NGC4649, could be ascribed to the effects of ram pressure
ablation.

\begin{figure}[h]

\vspace{8.pc}

\caption{\scriptsize Temperature profile of NGC4649, NGC5044 and NGC5846.
The ROSAT PSPC points are represented by dotted line crosses and show an
emission weighted temperature. Solid line crosses denote the joint ASCA and
ROSAT results, except for NGC5044, where they denote ASCA results.  Errors
are shown on the 1$\sigma$ confidence level. 1\amin\ corresponds to
4.94~kpc, 15.7~kpc, 9.74~kpc for NGC4649, NGC5044 and NGC5846,
respectively.}

\end{figure}

\begin{figure}[h]

\vspace{8.pc}

\caption{\scriptsize  Abundance profiles. Data points are represented as in
Fig.1. Solid line represents best fit abundance gradient in stars from
optical measurements and dashed lines indicate the acceptable range.
Central values for elemental abundance were taken as 2.3 for Si\&S and 1.4
for Fe for NGC5846 and 3.6 for Si\&S and 2.2 for Fe for NGC4649, following
measurements of FFI95 and theoretical calculations by MT87.}

\end{figure}

\vspace*{-1pc}

\section{References}


\re
1.      Anders E. and Grevesse N., 1989, {\it Geochimica et Cosmochimica Acta},
{\bf 53}, 197. \hfill AG89.
\re
2.      Carollo, C., M., Danziger, I., J. and Buson, L., 1993, {\it MNRAS},
{\bf 265}, 553. \hfill CDB93.
\re
3.      Ciotti, L. \etal, 1991, {\it Ap. J.}, {\bf
376}, 380. \hfill C91.
\re
4.      David L. \etal, 1994, {\it Ap. J.}, {\bf 428}, 544. \hfill D94.
\re
5. 	Finoguenov, A. Forman, W. and Jones, C., 1996, {\it Ap. J.},
submitted. \hfill FFJ96.
\re
6.      Fisher D., Franx M. and Illingworth G., 1995, {\it Ap. J.}, {\bf
448}, 119. \hfill FFI95.
\re
7.      Matteuci \& Tornamb\'e, 1987, {\it A\&A}, {\bf 185}, 51. \hfill MT87.
\re
8.     Woosley, S., E., and Weaver, T., A., 1986, {\it ARAA}, {\bf 24},
205. \hfill WW86.

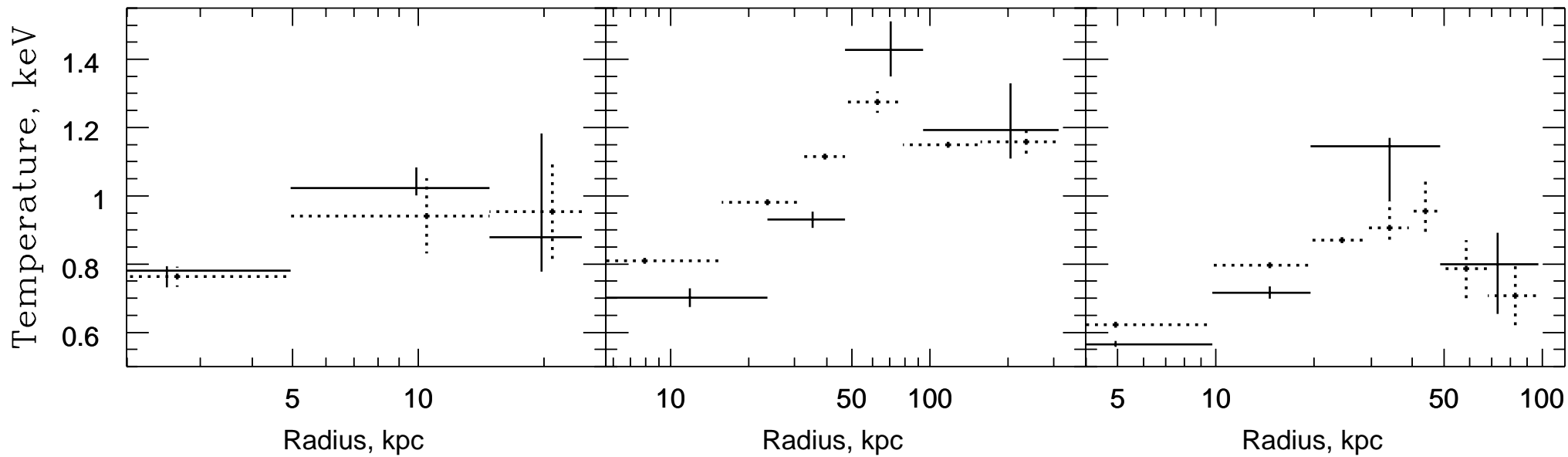
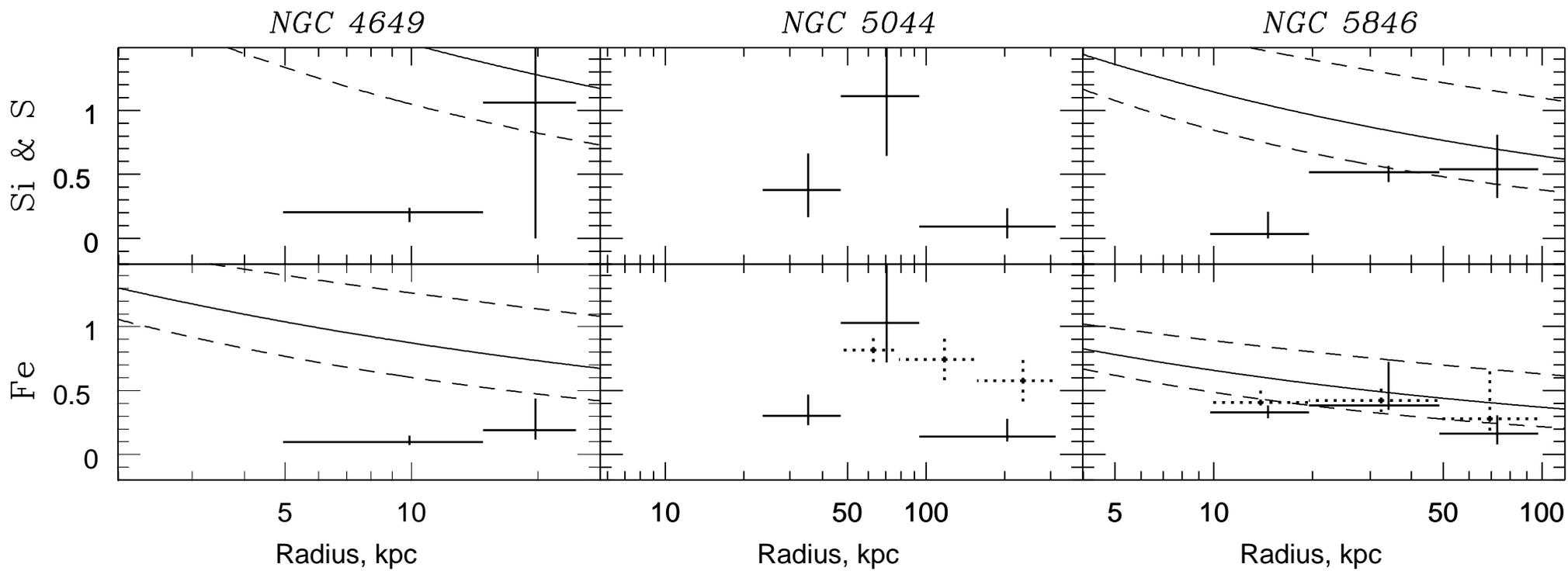
\end{document}